\documentclass[a4paper,12pt]{article}

\usepackage{amssymb}
\usepackage{mathrsfs}
\usepackage{bbm}

\newlength{\dinwidth}
\newlength{\dinmargin}
\newcommand{\qsq}{q^2}

\begin{document}

\title{\bf Constraints on $B\to \pi, K$ transition form factors from exclusive semileptonic D-meson decays}\bigskip

\author{Fang Su$^{1}$ and Ya-Dong Yang$^{1,2}$\\
{ $^1$\small Institute of Particle Physics, Huazhong Normal University, Wuhan, Hubei  430079, P.~R. China}\\
{ $^2$\small Key Laboratory of Quark \& Lepton Physics, Ministry of Education, Huazhong Normal University,}\\ {\small Wuhan, Hubei, 430079, P.~R. China}}

\date{}

\maketitle
\bigskip\bigskip
\maketitle
\vspace{-1.5cm}

\begin{abstract}

{\noindent}According to the heavy-quark flavour symmetry, the $B\to \pi, K$ transition form factors could be related to the corresponding ones of D-meson decays near the zero recoil point. With the recent precisely measured exclusive semileptonic decays $D \to \pi \ell \nu$ and $D\to K \ell \nu$, we perform a phenomenological study of $B \to \pi, K$ transition form factors based on this symmetry. Using BK, BZ and Series Expansion parameterizations of the form factor slope, we extrapolate $B \to \pi, K$ transition form factors from $q^{2}_{max}$ to $q^{2}=0$. It is found that, although being consistent with each other within error bars, the central values of our results for $B \to \pi, K$ form factors at $q^2=0$, $f_+^{B\to \pi, K}(0)$, are much smaller than predictions of the QCD light-cone sum rules, but are in good agreements with the ones extracted from hadronic B-meson decays within the SCET framework. Moreover, smaller form factors are also favored by the QCD factorization approach for hadronic B-meson decays.

\end{abstract}

\newpage

\section{Introduction}
\label{sec:intro}

Exclusive semileptonic B- and D-meson decays, such as $B\to \pi \ell \nu$, $D\to \pi\ell \nu$ and $D\to K \ell \nu$, provide a means both to measure fundamental Standard Model~(SM) parameters and to perform detailed studies of decay dynamics~\cite{Antonelli:2009ws,Asner:2008nq,Buchalla:2008jp,Richman:1995wm}. For example, one can determine the Cabibbo-Kobayashi-Maskawa~(CKM)~\cite{CKM} matrix elements $|V_{cs}|$ from $D\to K\ell\nu$, $|V_{cd}|$ from $D\to\pi\ell\nu$, and $|V_{ub}|$ from $B\to\pi\ell\nu$ through measurements of the corresponding differential decay rate. In order to extract precise information about these parameters, it is essential to have a good theoretical control over the hadronic matrix elements between the initial- and the final-state mesons, which are usually parameterized as transition form factors. This is, unfortunately, a very difficult task because of the involved non-perturbative hadronic effects.

There have been many studies devoted to the computation of these heavy-to-light transition form factors. Various quark models have been employed~(see, e.g., \cite{ISGW2,quark-model,light-front}), which in many aspects help our phenomenological understandings of the heavy-to-light transitions. More quantitative
predictions, however, require techniques based on first principles of QCD, such as lattice QCD~(LQCD) simulation~(see, e.g., \cite{LQCD-D,Dalgic:2006dt,Bailey:2008wp,Bernard:2009ke,LQCD-D-B}) and/or QCD sum rules on the light-cone~(LCSR)~(see, e.g., \cite{LCSR-D,Ball:2004ye,Duplancic:2008ix,Duplancic:2008tk,LCSR-B,LCSR-review}). These two methods are complementary to each other with respect to the momentum transfer $q^2$ between the initial- and final-state mesons: while the LQCD calculations are restricted to high  $q^2$ region, reliable predictions of the LCSR method can only be made at low  $q^2$ region. Neither the LQCD nor the LCSR calculations predict the form factors over the full $q^2$ region.

In order to interpolate between the results for small and large momentum transfers, various form factor parameterizations have been suggested in the literature: the two-parameter Be\'cirevi\'c-Kaidalov~(BK) ansatz~\cite{Becirevic:1999kt}, the three-parameter Ball-Zwicky~(BZ) ansatz~\cite{Ball:2004ye}, the so-called Series Expansion~(SE) representation~\cite{Arnesen:2005ez,Becher:2005bg,SE}, as well as the representation from the Omnes solution to the dispersive bounds~\cite{Omnes}. A good review of these different Ans\"{a}tze could be found in Refs.~\cite{Dalgic:2006dt,Besson:2009uv}.

In the past few years, much progress has been made in our understanding of B- and D-meson semileptonic decays. Especially for the latter, measurements have advanced in accuracy from $6\%$-$20\%$~\cite{D-data} on the normalization and $\sim 10\%$~\cite{Link:2004dh} on the shape to $\sim 1\%$ on both~\cite{Besson:2009uv} of the transition form factors. In an effective theory where both the charm and bottom quarks are treated as heavy, there is an SU(2)-flavour symmetry~\cite{Isgur:1989vq,Georgi:1990um,Neubert:1993mb}, which relates heavy quarks of the same four-velocity $v$ but different mass. Using this heavy-quark flavour symmetry~(HQFS), we can relate the $B\to \pi, K$ transition form factors to the precisely measured $D\to \pi, K$ ones near the zero-recoil point~\cite{Isgur:1990kf}, where $1/m_{Q}$ power corrections are estimated to be small~\cite{Burdman:1993es}. Combining the fit of the shape parameters from experimental data and/or other theoretical calculations with the HQFS relation, we can then extract the $B\to \pi, K$ form factors at zero momentum transfer.

Our paper is organized as follows. In Section~2, we provide the definition of heavy-to-light form factors and their different parameterizations, as well as the relationship between the B- and D-meson form factors based on the HQFS. In Section~3, after collecting the up-to-date measured $D\to \pi, K$ form-factor shape parameters, we use the HQFS to get the corresponding $B\to \pi, K$ form factors at the zero-recoil point; using the fitted results for the shape parameters in various parameterizations, we then give our final numerical results for $B\to \pi, K$ form factors at $q^2=0$; some interesting phenomenological discussions are also presented in this section. Our conclusions are made in Section~4.

\section{Heavy-to-light transition form factors}
\label{sec:theo}

\subsection{Definition of form factors}
\label{sec:FF-def}

The hadronic matrix elements for a generic semileptonic decay $H \to P \ell \nu$, where $H$ and $P$ denote a heavy and a light pseudoscalar meson respectively, is usually written in terms of two form factors $f_+(q^2)$ and $f_0(q^2)$~\cite{Ball:2004ye,Wirbel:1985ji},
\begin{equation}\label{eq:ffdef1}
    \langle P(p_P) | V^\mu | H(p_H) \rangle  = f_+(q^2)
	\left(p^\mu_H + p^\mu_P - \frac{m_H^2 - m_P^2}{q^2}\,q^\mu \right) +
	f_0(q^2) \frac{m_H^2 - m_P^2}{q^2}\,q^\mu ,
\end{equation}
where $q \equiv p_H - p_P$ is the momentum transferred to the lepton pair, and $V^\mu\equiv {\bar q}\gamma^\mu Q$ denotes the heavy-to-light vector current. The additional relation $f_+(0) = f_0(0)$ holds at the maximal recoil point.

In the case of massless leptons, which is a good approximation for $\ell=e$, $\mu$, the form factor $f_0(q^2)$ is absent and the differential decay rate for $H \to P \ell \nu$ involve $f_{+}(q^{2})$ only:
\begin{equation}\label{eq:slep}
    \frac{d\Gamma}{dq^2} = \frac{G_F^2 |V_{qQ}|^2}{192 \pi^3 m_H^3}
    \lambda^{3/2}(q^2) |f_+(q^2)|^2 ,
\end{equation}
with $\lambda(q^2)=(m_H^2 + m_P^2 - q^2)^2 - 4 m_H^2 m_P^2$, being the usual triangle function. To obtain
the total width, one should integrate Eq.~(\ref{eq:slep}) over the entire physical region, $0\leq q^2 \leq (m_H-m_P)^2$, which requires the precise knowledge of the normalization and the $q^2$-dependence of the form factor $f_+(q^2)$. On the other hand, a precise experimental measurement of the decay rate, in combination with a controlled theoretical calculation of the form factor, allows for a clean determination of the CKM matrix element $|V_{qQ}|$.

The heavy-to-light form factors are also needed as ingredients in the analysis of two-body hadronic B-meson decays, e.g., $B\to \pi K$, in the framework of QCD factorization~\cite{Beneke:1999br}, again with the objective to provide precision determinations of quark-flavor mixing parameters in the SM.

\subsection{Form-factor parameterizations}
\label{sec:FF-para}

Due to their special role in flavour physics, a variety of theoretical calculations for the heavy-to-light form factors exist in the literature. In additional to various quark models~(see, e.g., \cite{ISGW2,quark-model,light-front}), the current state-of-the-art methods based on first QCD principles  are LQCD simulations~(see, e.g., \cite{LQCD-D,Bailey:2008wp,Dalgic:2006dt,Bernard:2009ke,LQCD-D-B}) and QCD sum rules on the light-cone~(LCSR)~(see, e.g., \cite{LCSR-D,Ball:2004ye,Duplancic:2008ix,Duplancic:2008tk,LCSR-B,LCSR-review}). The latter two result in predictions for different $q^2$ regions. The LQCD calculations are only available in the high-$q^2$ region, while LCSR provides information near $q^2=0$; neither of them can predict the form factors over the full $q^2$ range. Interpolations between these two regions are therefore needed.

While the prediction of the exact form-factor shape is challenged for any theoretical calculations, it is well established that the general properties of analyticity, cross symmetry, and unitarity largely constrain the possible behavior of the form factor in the semileptonic region~\cite{SE,Bourrely:1980gp,Lellouch:1995yv}. Specifically, it is expected to be an analytic function everywhere in the complex $q^2$ plane outside of a cut that extends along the positive $q^2$ axis from the mass of the lowest-lying $b(c)\bar{d}$~(for $B(D)\rightarrow\pi$) or $b(c)\bar{s}$~(for $B(D)\rightarrow K$) vector meson. This assumption leads to a dispersion relation~(see, e.g., \cite{Becher:2005bg,Hill:2005ju}):
\begin{equation}\label{eq:dispersion}
   f_+\left(q^2\right) = \frac{f_+\left(0\right)/(1-\alpha)}{1-\frac{\qsq}{M^2_{H^*_{(s)}}}} + \frac{1}{\pi}\int_{(m_H+m_P)^2}^\infty {\frac{{\rm Im}f_+\left(t\right)}{t-\qsq-i\epsilon}dt},
\end{equation}
where $m_{H^*_{(s)}}$ is the mass of meson $H^*$ for $H\to \pi$ or $H^*_s$ for $H\to K$, and $\alpha$ gives the relative size of contribution to $f_+(0)$ from the $H^*_{(s)}$ meson. Most of the suggested form factor parameterizations in the literature are motivated by a version of the dispersion relation Eq.~(\ref{eq:dispersion}), with the integral being approximated by a number of effective poles~\cite{Becher:2005bg,Hill:2005ju},
\begin{equation}\label{eq:disp_rel_sum}
   f_+\left(q^2\right) = \frac{f_+\left(0\right)/(1-\alpha)}{1-\frac{\qsq}{M^2_{H^*_{(s)}}}} +\sum_{k=1}^{N}{\frac{\rho_k}{1-\frac{1}{\gamma_k}\frac{q^2}{M_{H^*_{(s)}}^2}}},
\end{equation}
where the expansion parameters $\rho_k$ and $\gamma_k$ are unknown.

A parametrization known as the simple pole model~(SP) assumes that the sum in Eq.~(\ref{eq:disp_rel_sum}) is dominated by a single pole~\cite{Becirevic:1999kt},
\begin{equation}\label{eq:SP-ansatz}
f_+(q^2) = \frac{f_+(0)}{1-q^2/m_{\rm pole}^2},
\end{equation}
where $f_+(0)$ sets the normalization of the form factor, and the value of $m_{\rm pole}$ is taken to be $M_{D^*_{(s)}}$ for $D\to \pi(K)$ or $M_{B^*_{(s)}}$ for $B\to \pi(K)$. The SP ansatz is certainly too naive, since there is no reason for the lowest-lying pole to saturate the form factor.

Another parametrization, known as the modified pole model~(MP) or Be\'cirevi\'c-Kaidalov~(BK) ansatz~\cite{Becirevic:1999kt}, adds a second term to the expansion given in Eq.~(\ref{eq:disp_rel_sum}), thus assuming that all higher-order poles can be modeled by a single effective pole. After making some further simplifying assumptions, the two pole terms are reduced to~\cite{Becirevic:1999kt}
\begin{equation}\label{eq:BK-ansatz}
f_+(q^2) = \frac{f_+(0)}{(1-q^2/m_{\rm pole}^2)(1-\alpha_{BK}\,q^2/m_{\rm pole}^2)},
\end{equation}
where $m_{\rm pole}$ is generally fixed to the $H^*_{(s)}$ mass, $f_+(0)$ sets the normalization and $\alpha_{BK}$ defines the shape of the form factor. As an extension of the BK ansatz, Ball and Zwicky proposed to use the following parametrization~(the so-called Ball-Zwicky~(BZ) ansatz)~\cite{Ball:2004ye},
\begin{equation}\label{eq:BZ-ansatz}
f_+(q^2) = f_+(0)\left[\frac{1}{1-q^2/m_{\rm pole}^2}+\frac{r_{BZ}\,q^2/m_{\rm pole}^2}{(1-q^2/m_{\rm pole}^2)(1-\alpha_{BZ}\,q^2/m_{\rm pole}^2)}\right],
\end{equation}
with the two shape parameters $\alpha_{BZ}$ and $r_{BZ}$, and the normalization $f_+(0)$. It is related to the BK ansatz by the simplification $\alpha_{BK}=\alpha_{BZ}=r_{BZ}$. Both the BK and BZ Ans\"{a}tze incorporate many of the known properties of the form factor, such as the kinematic constraint at $q^2=0$, heavy-quark scaling, and the location of the lowest-lying pole. They are also featured by both being intuitive and having fewer free parameters.

Although these parameterizations have been widely adopted in the literature, the presence of poles near the semileptonic $q^2$ regions causes the sum in Eq.~(\ref{eq:disp_rel_sum}) to have poor convergence properties, creating doubt as to whether truncating all but the first one or two terms leaves an accurate estimate of the true form-factor shape. Another class of parameterizations, known as the Series Expansion~(SE)~\cite{Arnesen:2005ez,Becher:2005bg,SE}, attempts to address the problem~\cite{Becher:2005bg,Hill:2006ub}.

Exploiting the analytic properties of $f_+(q^2)$, a transformation of variables is made that maps $q^2$ in the semileptonic region onto a unit circle $|z|<1$, where
\begin{equation}\label{eq:z-para}
  z(q^2,t_0) = \frac{\sqrt{t_+ -\qsq}-\sqrt{t_+ -t_0}}{\sqrt{t_+ -q^2}+\sqrt{t_+ -t_0}},
\end{equation}
$t_\pm=(m_H \pm m_P)^2$, and $t_0$ is any real number less than $t_+$. In terms of this new variable, $z(q^2,t_0)$, the SE ansatz corresponds to the following form-factor parametrization~\cite{Arnesen:2005ez,Becher:2005bg,SE,Hill:2006ub}:
\begin{equation}\label{eq:SE-ansatz}
f_+(q^2) = \frac{1}{P(q^2)\phi(q^2,t_0)}\sum_{k=0}^\infty a_k\left(t_0\right)\left[z\left(q^2,t_0\right)\right]^k,
\end{equation}
The Blaschke factor $P(q^2)$ accounts for the low-lying resonances presented below $t_+$. Since $m_{D^*}>m_{D}+m_{\pi}$, while $m_{D^*_s}<m_{D}+m_{K}$, $m_{B^*}<m_{B}+m_{\pi}$, and $m_{B^*_s}<m_{B}+m_{K}$, we therefore have
\begin{eqnarray}
    P^{D\to\pi\ell\nu}(q^2)  = 1, && \qquad
    P^{D\to K \ell\nu}(q^2)  = z(q^2, m_{D^*_s}^2), \\
    P^{B\to\pi\ell\nu}(q^2)  = z(q^2, m_{B^*}^2), && \qquad
    P^{B\to K \ell\nu}(q^2)  = z(q^2, m_{B^*_s}^2).
\end{eqnarray}
In Eq.~(\ref{eq:SE-ansatz}), $\phi(q^2, t_0)$ is an arbitrary analytic function outside a cut in the complex $q^2$ plane from $t_+$ to $\infty$. The standard choice for $\phi(q^2,t_0)$ is~\cite{Arnesen:2005ez}
\begin{eqnarray}
    \phi_+(q^2, t_0)  &=&  \sqrt{\frac{1}{32 \pi \chi^{(0)}_J}}
    \left( \sqrt{t_+ - q^2} + \sqrt{t_+ - t_0}  \right)
	\left( \sqrt{t_+ - q^2} + \sqrt{t_+ - t_-} \right)^{3/2}  \nonumber \\
    & \times & \left( \sqrt{t_+ - q^2} + \sqrt{t_+} \right)^{-5}
	\frac{(t_+ - q^2)}{(t_+ - t_0)^{1/4}},
    \label{eq:slep:phi}
\end{eqnarray}
where the factor $\chi^{(0)}_J$ can be calculated using perturbation theory and the operator product expansion~\cite{Arnesen:2005ez,Bharucha:2010im}. A variant form for $\phi(q^2,t_0)$ could be found in Ref.~\cite{Becher:2005bg}.

In order to accelerate the convergence of the power series in $z$, the free parameter $t_0$ in Eq.~(\ref{eq:z-para}) can be chosen to make the range of $|z|$ as small as possible. With the traditional choice $t_0=t_+(1-\sqrt{1-t_-/t_+})$~\cite{Becher:2005bg}, as well as the above choices for $P(q^2)$ and $\phi(q^2, t_0)$, it can be shown that the sum over all k of the series coefficients $a_k^2$ is of order unity by unitarity bound or even much less than one due to the heavy-quark bound~\cite{Becher:2005bg}. The tight heavy-quark constraint on the size of the coefficients in the $z$-expansion, in conjunction with the small value of $|z|$, ensures that only the first few terms in the series are needed to describe
heavy-to-light semileptonic form factors to a high accuracy. Thus, the SE ansatz has improved convergence properties over Eq.~(\ref{eq:disp_rel_sum}).

\subsection{Form-factor relationship based on heavy-quark flavour symmetry}
\label{sec:FF-relation}

The heavy-quark flavour symmetry~(HQFS) relations for form factors were first obtained by Isgur and Wise in a different parameterization for the hadronic matrix elements~\cite{Isgur:1990kf},
\begin{equation}\label{eq:ffdef2}
    \langle P(p_P) | V^\mu | H(p_H) \rangle = f_+(q^2) \left(p^\mu_H + p^\mu_P\right) + f_-(q^2) \left(p^\mu_H - p^\mu_P\right),
\end{equation}
where the form factor $f_-(q^2)$ is related to $f_+(q^2)$ and $f_0(q^2)$ defined in Eq.~(\ref{eq:ffdef1}) by \begin{equation}
    f_-(q^2) = \frac{m_H^2-m_P^2}{q^2}\left[f_0(q^2)-f_+(q^2)\right].
\end{equation}
Based on the HQFS symmetry, it follows that~\cite{Isgur:1990kf}:
\begin{eqnarray}\label{eq:FF-relations}
    (f_{+}+f_{-})^{B\to L} &=& \sqrt{\frac{m_D}{m_B}}\;\left[\frac{\alpha_s(m_b)}
    {\alpha_s(m_c)}\right]^{-6/25}\;(f_{+}+f_{-})^{D \to L} \nonumber \\
    (f_{+}-f_{-})^{B\to L} &=& \sqrt{\frac{m_B}{m_D}}\;\left[\frac{\alpha_s(m_b)}
    {\alpha_s(m_c)}\right]^{-6/25}\;(f_{+}-f_{-})^{D \to L},
\end{eqnarray}
where $L$ denotes a light pseudoscalar meson like $\pi$ or $K$. As emphasized in Ref.~\cite{Isgur:1990kf}, these relations certainly hold near the zero recoil point where $q^2$ is near its maximum value $q_{max,H}^2=t_-=(m_H-m_L)^2$. Furthermore, at this kinematic limit, the form factors $f_{\pm}(t_-)$ obey the so-called Isgur-Wise scaling laws~\cite{Isgur:1989qw},
\begin{equation}\label{eq:FF-scaling}
    f_{+}(t_-)+f_{-}(t_-) \sim m_Q^{-1/2}, \qquad f_{+}(t_-)-f_{-}(t_-) \sim m_Q^{+1/2},
\end{equation}
where $m_Q$ denotes the heavy-quark mass, and logarithms of $m_Q$ arising from perturbative QCD corrections have been neglected. To the leading order in $1/m_Q$, we can therefore set $f_{-}(t_-) \simeq -f_{+}(t_-)$. With this approximation and relations Eq.~(\ref{eq:FF-relations}), we finally get
\begin{equation}\label{eq:FF-relation}
    f_{+}^{B\to L}(q_{max,B}^2) = \sqrt{\frac{m_B}{m_D}}\;\left[\frac{\alpha_s(m_b)}{\alpha_s(m_c)}\right]^{-6/25}\;f_{+}^{D \to L}(q_{max,D}^2).
\end{equation}

The reliability of the relation Eq.~(\ref{eq:FF-relation}) depends on the importance of symmetry-breaking corrections of order $1/m_Q$. Fortunately, as discussed in detail by Burdman et al.~\cite{Burdman:1993es}, the power corrections to $f_+(q^2)$ turn out to be small, being only of order $15\%$~\cite{Burdman:1993es}. Therefore, in principle, we can evaluate the $B\to P$ transition form factors from the corresponding ones in D-meson semileptonic decays, which have been recently measured to a new level of precision by the CLEO~\cite{Besson:2009uv}, Belle~\cite{Widhalm:2006wz}, and BaBar~\cite{Aubert:2007wg} collaborations.

This idea has been used to estimate the CKM matrix element $|V_{ub}|$~\cite{Ligeti:1997aq}, to study the rare $B\to V \gamma$ and $B\to V \ell^+ \ell^-$ decays~\cite{Isgur:1990kf,Ligeti:1999th}, as well as to study two-body hadronic B-meson decays~\cite{Buccella:1993nm} and so on~\cite{Falcone:1998gs}.

\section{Numerical results and discussions}
\label{sec:results}

Combining the form factor parameterizations and the HQFS relation with the precisely measured $D\to \pi \ell \nu$ and $D\to K \ell \nu$ decays~\cite{Besson:2009uv,Widhalm:2006wz,Aubert:2007wg}, it is possible to extract the corresponding $B\to \pi$ and $B\to K$ form factors, which will be detailed in this section.

\subsection{Measured $D\to \pi, K$ transition form factors}
\label{sec:FF-D}

In the last few years, a new level of precision has been achieved in measurements of branching fractions of exclusive semileptonic decays $D\to \pi \ell \nu$ and $D\to K \ell \nu$ by the CLEO~\cite{Besson:2009uv}, Belle~\cite{Widhalm:2006wz}, and BaBar~\cite{Aubert:2007wg} collaborations, using complementary experimental approaches. These results are all highly consistent with each other. For more details, the readers are referred to these original references.

These experiments have also performed studies of the $q^2$ parameterizations of the form factors and extracted the associated shape parameters. A summary of the form factor parameters obtained for isospin-combined $D\to \pi \ell \nu$ and $D\to K \ell \nu$ decays is given in Tables~\ref{tab:FFDtopisumary} and \ref{tab:FFDtoKsumary}, respectively.

\begin{table}[t]
\begin{center}
\caption{\label{tab:FFDtopisumary} \small Summary of the form-factor parameters obtained by different experiments for isospin-combined $D\to \pi \ell \nu$ decay, as well as the values obtained by LQCD simulation~\cite{Bernard:2009ke}. The uncertainties are statistical, systematic and, where applicable, from external inputs like CKM elements. The former two on the least significant digits are shown in parentheses for SE ansatz.}
\vspace{0.1cm}
\begin{tabular}{llll}
\hline \hline
SP and BK:  & \multicolumn{1}{c}{$m_{\rm pole}[{\rm GeV}]$} & \multicolumn{1}{c}{$\alpha_{BK}$}
& \multicolumn{1}{c}{$f_+(0)$} \\
\hline
CLEO~\cite{Besson:2009uv}   & $1.91 \pm 0.02 \pm 0.01$ & $0.21 \pm 0.07 \pm 0.02$ & $0.666 \pm 0.019 \pm 0.004 \pm 0.003$\\
Belle~\cite{Widhalm:2006wz} & $1.97 \pm 0.08 \pm 0.04$ & $0.10 \pm 0.21 \pm 0.10$ & $0.624 \pm 0.020 \pm 0.030$ \\
LQCD~\cite{Bernard:2009ke}  &                          & $0.44 \pm 0.04 \pm 0.07$ & $0.64  \pm 0.03  \pm 0.06$ \\
\hline
3 para. SE: & \multicolumn{1}{c}{$a_0$}                     & \multicolumn{1}{c}{$a_1$}
& \multicolumn{1}{c}{$a_2$} \\
\hline
CLEO~\cite{Besson:2009uv}   & $\quad 0.072(2)(1)$ & $\quad -0.17(3)(1)$ & $\qquad \qquad 0.3(2)(0)$ \\
\hline
2 para. SE: & \multicolumn{1}{c}{$a_0$}                     & \multicolumn{1}{c}{$a_1$}
& \\
\hline
CLEO~\cite{Besson:2009uv}   & $\quad 0.071(2)(1)$ & $\quad -0.13(1)(0)$ & \\
\hline \hline
\end{tabular}
\end{center}
\end{table}

\begin{table}[t]
\begin{center}
\caption{\label{tab:FFDtoKsumary} \small The same as Table~\ref{tab:FFDtopisumary} but for isospin-combined $D\to K \ell \nu$ decay.}
\vspace{0.1cm}
\begin{tabular}{llll}
\hline \hline
SP and BK:  & \multicolumn{1}{c}{$m_{\rm pole}[{\rm GeV}]$} & \multicolumn{1}{c}{$\alpha_{BK}$}
& \multicolumn{1}{c}{$f_+(0)$} \\
\hline
CLEO~\cite{Besson:2009uv}   & $1.93 \pm 0.02 \pm 0.01$    & $0.30 \pm 0.03 \pm 0.01$ & $0.739 \pm 0.007 \pm 0.005$\\
Belle~\cite{Widhalm:2006wz} & $1.82 \pm 0.04 \pm 0.03$    & $0.52 \pm 0.08 \pm 0.06$ & $0.695 \pm 0.007 \pm 0.022$ \\
BaBar~\cite{Aubert:2007wg}  & $1.884 \pm 0.012 \pm 0.015$ & $0.38 \pm 0.02 \pm 0.03$ & $0.727 \pm 0.007 \pm 0.005 \pm 0.007$ \\
LQCD~\cite{Bernard:2009ke}  &                             & $0.50 \pm 0.04 \pm 0.07$ & $0.73  \pm 0.03  \pm 0.07$ \\
\hline
3 para. SE: & \multicolumn{1}{c}{$a_0$}                     & \multicolumn{1}{c}{$a_1$}
& \multicolumn{1}{c}{$a_2$} \\
\hline
CLEO~\cite{Besson:2009uv}   & $\quad 0.0263(1)(2)$ & $\quad -0.06(1)(0)$ & $\qquad \qquad 0.1(2)(0)$ \\
\hline
2 para. SE: & \multicolumn{1}{c}{$a_0$}                     & \multicolumn{1}{c}{$a_1$}
& \\
\hline
CLEO~\cite{Besson:2009uv}   & $\quad 0.0263(1)(2)$ & $\quad -0.056(4)(2)$ & \\
\hline \hline
\end{tabular}
\end{center}
\end{table}

As concluded in Ref.~\cite{Besson:2009uv}, the quantity of the fits is good for all these parameterizations: the two- and three-parameter~(2 para. and 3 para.) SE, as well as the BK Ans\"{a}tze. The fitted results for $m_{\rm pole}$ in the SP ansatz show, however, some deviations from the predicted values~(especially for $m_{D_s^*}$), indicating that the lowest-lying pole could not saturate the form factor. The fitted results within the BK ansatz are also consistent with the recent LQCD calculation~\cite{LQCD-D,Bernard:2009ke}; especially as show in Figure~2 in Ref.~\cite{Bernard:2009ke}, the agreement between the experimental measurements and the LQCD computation is good for $D\to \pi \ell \nu$ and very good for $D\to K \ell \nu$.

\subsection{$B\to \pi, K$ form factors based on HQFS}
\label{sec:FF-B}

Based on the HQFS relation Eq.~(\ref{eq:FF-relation}) and the measured $D\to \pi, K$ transition form factors presented in Tables~\ref{tab:FFDtopisumary} and \ref{tab:FFDtoKsumary}, we can now get the corresponding ones of $B\to \pi, K$ decays at the zero recoil point.

Before presenting the results for $f_{+}^{B\to \pi, K}(q_{max,B}^2)$, we would like to first fix the relevant input parameters, such as the quark and meson masses and the running coupling constant. For the charm- and the bottom-quark masses, we adopt the values determined from vector-current correlators and experimental $e^+e^-$ collisions with ${\cal O}(\alpha_s^3)$ accuracy~\cite{b-c-mass},
\begin{equation}\label{eq:b-c-mass}
\bar{m}_b(\bar{m}_b)=4.163~{\rm GeV}\,, \qquad \bar{m}_c(\bar{m}_c)=1.279~{\rm GeV}\,.
\end{equation}
These results are in good agreement with the recent lattice determination~\cite{McNeile:2010ji}. For all the other parameters, we list them in Table~\ref{tab:Inputs}. Throughout the paper, we use the isospin-averaged meson masses, for example, $m_{\pi}=(m_{\pi^+}+m_{\pi^0})/2$. To run the QCD coupling constant from the initial scale $m_Z$ down to the lower scales $m_b$ and $m_c$, we have used the Mathematica package RunDec~\cite{Chetyrkin:2000yt}.

\begin{table}[t]
\begin{center}
\caption{\label{tab:Inputs} \small The relevant input parameters used in our calculation. All meson masses are taken directly from Particle Data Group~\cite{Amsler:2008zzb}.}
\vspace{0.1cm}
\begin{tabular}{lllll}
\hline \hline
$m_{\pi^+}=139.6~{\rm MeV}$ & $m_{\pi^0}=135.0~{\rm MeV}$ & $m_{K^+}=493.7~{\rm MeV}$ &
$m_{K^0}=497.6~{\rm MeV}$ \\
\hline
$m_{B^+}=5279.2~{\rm MeV}$ & $m_{B^0}=5279.5~{\rm MeV}$ & $m_{D^+}=1869.6~{\rm MeV}$ &
$m_{D^0}=1864.8~{\rm MeV}$ \\
\hline
$m_{B^{\ast}}=5325.1~{\rm MeV}$ & $m_{B^{\ast}_s}=5415.4~{\rm MeV}$ & $m_{D^{\ast}}=2007.0~{\rm MeV}$ &
$m_{D^{\ast}_s}=2112.3~{\rm MeV}$ \\
\hline
$\alpha_s(m_Z)=0.1184$~\cite{Bethke:2009jm} & $m_{Z}=91.188~{\rm GeV}$~\cite{Amsler:2008zzb} & & \\
\hline \hline
\end{tabular}
\end{center}
\end{table}

With all the above equipments, our final numerical results for $f_{+}^{B\to \pi}(q_{max,B}^2)$ with $q_{max,B}^2=(m_B-m_{\pi})^2=26.44~{\rm GeV}^2$, and $f_{+}^{B\to K}(q_{max,B}^2)$ with $q_{max,B}^2=(m_B-m_{K})^2=22.88~{\rm GeV}^2$ are given, respectively, in the first row of Tables~\ref{tab:FFBtopi1} and \ref{tab:FFBtoK1}, where the uncertainties come mainly from the shape parameters. These results could be used to test predictions based on the effective heavy meson chiral lagrangian, describing the interactions of low-momentum pions with mesons containing a single heavy quark~\cite{Wise:1992hn}~(for a review, see \cite{Casalbuoni:1996pg}). For example, a larger value, $f_{+}^{B\to \pi}(26.42)=10.38 \pm 3.63$ is quoted by Arnesen et al.~\cite{Arnesen:2005ez}. With their respective uncertainties taken into account, our results are also consistent with the LQCD simulation, $f_{+}^{B\to \pi}(26.5)=9.04 \pm 2.21 \pm 0.53$, where the first error is statistical plus the one from chiral perturbative theory, and the second systematic~\cite{Bailey:2008wp}.

\begin{table}[t]
\begin{center}
\caption{\label{tab:FFBtopi1} \small Numerical results for $f_{+}^{B\to \pi}(q_{max,B}^2)$~(the first row in each case) based on the HQFS relation and the measured $D\to \pi$ form-factor shape parameters listed in Table~\ref{tab:FFDtopisumary}, as well as the normalization $f_{+}^{B\to \pi}(0)$ based on the fitted BK~(the second row) and BZ~(the third row) parameters Eq.~(\ref{eq:paraball})~\cite{Ball:2006jz}. The uncertainties come mainly from the form-factor shape parameters.}
\vspace{0.1cm}
\tabcolsep 0.2in
\begin{tabular}{lllll}
\hline \hline
based on
& \multicolumn{1}{c}{SP} & \multicolumn{1}{c}{BK} & \multicolumn{1}{c}{2 para. SE} & \multicolumn{1}{c}{3 para. SE} \\
\hline
CLEO~\cite{Besson:2009uv}   & $6.890^{+0.769}_{-0.628}$ & $5.706^{+0.412}_{-0.373}$
& $5.471^{+0.154}_{-0.154}$ & $6.432^{+0.465}_{-0.465}$ \\
                            & $0.235^{+0.037}_{-0.034}$ & $0.195^{+0.026}_{-0.025}$
& $0.187^{+0.021}_{-0.021}$ & $0.220^{+0.029}_{-0.029}$ \\
                            & $0.239^{+0.050}_{-0.043}$ & $0.198^{+0.037}_{-0.033}$
& $0.189^{+0.034}_{-0.030}$ & $0.223^{+0.042}_{-0.038}$ \\
\hline
Belle~\cite{Widhalm:2006wz} & $5.068^{+2.516}_{-1.170}$ & $4.875^{+1.139}_{-0.810}$ & & \\
                            & $0.173^{+0.088}_{-0.044}$ & $0.167^{+0.043}_{-0.033}$ & & \\
                            & $0.175^{+0.092}_{-0.049}$ & $0.169^{+0.049}_{-0.038}$ & & \\
\hline \hline
\end{tabular}
\end{center}
\end{table}

\begin{table}[t]
\begin{center}
\caption{\label{tab:FFBtoK1} \small Numerical results for $f_{+}^{B\to K}(q_{max,B}^2)$~(the first row in each case) based on the HQFS relation and the measured $D\to K$ form factor parameters listed in Table~\ref{tab:FFDtoKsumary}, as well as the normalization $f_{+}^{B\to K}(0)$~(the second row) based on the fitted shape parameters Eq.~(\ref{eq:a0b1range})~\cite{Bartsch:2009qp}. The uncertainties come mainly from the form-factor shape parameters.}
\vspace{0.1cm}
\tabcolsep 0.2in
\begin{tabular}{lllll}
\hline \hline
based on
& \multicolumn{1}{c}{SP} & \multicolumn{1}{c}{BK} & \multicolumn{1}{c}{2 para. SE} & \multicolumn{1}{c}{3 para. SE} \\
\hline
CLEO~\cite{Besson:2009uv}   & $2.775^{+0.070}_{-0.066}$ & $2.719^{+0.052}_{-0.051}$
& $2.665^{+0.026}_{-0.026}$ & $2.707^{+0.070}_{-0.070}$ \\ \vspace{0.2cm}
                            & $0.195^{+0.161}_{-0.036}$ & $0.191^{+0.157}_{-0.035}$
& $0.187^{+0.154}_{-0.035}$ & $0.190^{+0.157}_{-0.035}$ \\
\hline
Belle~\cite{Widhalm:2006wz} & $2.990^{+0.263}_{-0.218}$ & $2.861^{+0.189}_{-0.174}$ & & \\ \vspace{0.2cm}
                            & $0.210^{+0.174}_{-0.042}$ & $0.201^{+0.166}_{-0.039}$ & & \\
\hline
BaBar~\cite{Aubert:2007wg}  & $2.875^{+0.081}_{-0.077}$ & $2.782^{+0.071}_{-0.069}$ & & \\ \vspace{0.2cm}
                            & $0.202^{+0.167}_{-0.038}$ & $0.195^{+0.161}_{-0.036}$ & & \\
\hline \hline
\end{tabular}
\end{center}
\end{table}

\subsection{Numerical results for $f_+^{B\to \pi, K}(0)$}

The $B\to \pi$ and $B\to K$ transition form factors at zero momentum transfer, $f_+^{B\to \pi, K}(0)$, are also important ingredients in two-body hadronic B-meson decays within the framework of QCD factorization~\cite{Beneke:1999br}. In this subsection, assuming a definite behavior of the $q^2$ dependence and using the input $f_{+}(q_{max,B}^2)$ determined above, we shall extract the normalized form factors at $q^2=0$.

\subsubsection{\boldmath $f_+^{B\to \pi}(0)$:} For the $B\to \pi$ transition form factor $f_+(q^2)$, thanks to the precisely measured $q^2$ spectrum in semileptonic $B\to \pi \ell \nu$ decay~\cite{Aubert:2006px,:2010uj}, the form-factor shape parameters can be extracted with a good accuracy~\cite{Becher:2005bg,Aubert:2006px,:2010uj,Ball:2006jz}. Using the fitted shape parameters~\cite{Ball:2006jz},
\begin{equation}\label{eq:paraball}
\alpha_{BK}=0.53 \pm 0.06, \qquad \alpha_{BZ}=0.40^{+0.15}_{-0.22}, \qquad r_{BZ}=0.64^{+0.14}_{-0.13}\,,
\end{equation}
our final numerical results for $f_+^{B\to \pi}(0)$, corresponding to the different inputs $f_{+}^{B\to \pi}(q_{max,B}^2)$, are given in the second~(based on the fitted BK parameters) and the third~(based on the fitted BZ parameters) rows of Table~\ref{tab:FFBtopi1}, respectively. It is noted that both the BK and the BZ parameterizations for $B\to \pi$ form factor give almost the same values in each column.

Comparing our results listed in Table~\ref{tab:FFBtopi1} with the most recent LCSR calculations,
\begin{equation}\label{eq:LCSR-Btopi}
 f_+^{B\to \pi}(0) = \left\{\begin{array}{ll} \vspace{0.2cm}
 0.258 \pm 0.031~\cite{Ball:2004ye}, \\
 0.26^{+0.04}_{-0.03}~\cite{Duplancic:2008ix},
 \end{array}\right.
\end{equation}
we can see that, while the results using the CLEO data based on the SP and 3 para. SE Ans\"{a}tze are consistent with Eq.~(\ref{eq:LCSR-Btopi}) within their respective error bars, all the other sets, especially the ones from the Belle data, are consistently smaller than the LCSR predictions.

Through a $\chi^2$ fit to the current available experimental data on $B\to PP$ and $B\to VP$ decays within the framework of soft-collinear effective theory~(SCET)~\cite{SCET-theory} for hadronic B-meson decays~\cite{SCET-pheno}, Williamson and Zupan~\cite{Williamson:2006hb}, and Wang et al.~\cite{Wang:2008rk} have extracted the transition form factor,
\begin{equation}\label{eq:SCET-Btopi}
 f_+^{B\to \pi}(0) = \left\{\begin{array}{lll} \vspace{0.2cm}
 0.176 \pm 0.007~\cite{Williamson:2006hb}, \\ \vspace{0.2cm}
 0.192 \pm 0.005~[0.198 \pm 0.003]\, \qquad [{\rm solution 1}]~\cite{Wang:2008rk}, \\
 0.201 \pm 0.015~[0.206 \pm 0.004]\, \qquad [{\rm solution 2}]~\cite{Wang:2008rk},
 \end{array}\right.
\end{equation}
where the numbers in the brackets are results including the chirally-enhanced penguin contribution; for more details, we refer the readers to Ref.~\cite{Wang:2008rk}. It can be clearly seen that our results are in good agreements with these fits. In addition, in order to better describe the current experimental data on tree-dominated $B\to \pi\pi$ decays, the QCD factorization predictions based on the next-to-next-to-leading order calculation also prefer a smaller form factor~\cite{Beneke:2009ek}.

\subsubsection{\boldmath $f_+^{B\to K}(0)$:} In principle, both the shape and the normalization of the $B\to K$ transition form factor $f_+(q^2)$ could also be extracted from the rare $B\to K \ell^+ \ell^-$ and $B\to K \nu \bar\nu$ decays. It is, however, not feasible at present due to the lack of precise information on the dilepton invariant mass spectra. The situation should be improved in the near future, ultimately allowing for a precise determination of the form-factor shape parameters directly from experimental data~\cite{Antonelli:2009ws,Buchalla:2008jp}.

Following Ref.~\cite{Bartsch:2009qp}, we use the following parametrization for the form factor $f_+^{B\to K}(q^2)$,
\begin{equation}\label{eq:ffparam}
f_{+}^{B\to K}(s) = f_{+}^{B\to K}(0)\,\frac{1-(b_0+b_1-a_0 b_0)s}{(1-b_0 s)(1-b_1 s)}\,,
\end{equation}
where $s=q^2/m_B^2$, and $b_0=m^2_{B}/m^2_{B_{s}^{\ast}} \simeq 0.95$, represents the position of the $B_{s}^{\ast}$ pole and will be treated as fixed. The ranges for the remaining shape parameters $a_0$ and $b_1$ are
\begin{equation}\label{eq:a0b1range}
1.4\leq a_0\leq 1.8\,, \qquad 0.5\leq b_1/b_0\leq 1.0\,,
\end{equation}
with $a_0=1.6$ and $b_1/b_0=1.0$ as the default values, which are also consistent with the LCSR result~\cite{Ball:2004ye}.

Using the above information about the form-factor shape and parametrization, our final numerical results for $f_+^{B\to K}(0)$, corresponding to the different inputs $f_{+}^{B\to K}(q_{max,B}^2)$, are given in the second row of Table~\ref{tab:FFBtoK1}. Here the main uncertainty is due to the shape parameter $b_1$. Comparing with the recent LCSR predictions,
\begin{equation}\label{eq:LCSR-BtoK}
 f_+^{B\to K}(0) = \left\{\begin{array}{ll} \vspace{0.2cm}
 0.331 \pm 0.041~\cite{Ball:2004ye}, \\ \vspace{0.2cm}
 0.304 \pm 0.042~\cite{Bartsch:2009qp}, \\
 0.36^{+0.05}_{-0.04}~\cite{Duplancic:2008ix},
 \end{array}\right.
\end{equation}
we can see that, although being roughly consistent with each other with the large uncertainties taken into account, the default values of our results in Table~\ref{tab:FFBtoK1} are also much lower than the LCSR predictions. On the other hand, our default results are in good agreements with the ones extracted from hadronic B-meson decays within the SCET framework~\cite{Williamson:2006hb,Wang:2008rk}; assuming an SU(3)-flavour symmetry, we have $f_+^{B\to K}(0)=f_+^{B\to \pi}(0)$ with the latter given by Eq.~(\ref{eq:SCET-Btopi}).

\section{Conclusions}

In this paper, motivated by the much precisely measured semileptonic decays $D\to \pi \ell \nu$ and $D\to K \ell \nu$ by the CLEO~\cite{Besson:2009uv}, Belle~\cite{Widhalm:2006wz}, and BaBar~\cite{Aubert:2007wg} collaborations, we have performed a phenomenological study of the $B \to \pi, K$ transition form factors based on the heavy-quark flavour symmetry, which relates the former to the corresponding D-meson semileptonic form factors near the zero recoil point.

Through a detailed analysis, we found that, while being consistent with the information obtained from experimental data and theoretical calculations of B-meson decays within error bars, the central values of our results for $B \to \pi, K$ transition form factors are much smaller than predictions of light-cone QCD sum rules. However, our results are in good agreements with the ones extracted from hadronic B-meson decays within the SCET framework. In addition, smaller form factors are also favored by the QCD factorization approach for hadronic B-meson decays.

As remarked by Isgur and Wise~\cite{Isgur:1990kf}, our results, while less complete than a real QCD calculation like light-cone QCD sum rules and/or lattice QCD, are systematic consequences of QCD; the corrections to our results are suppressed by powers of $1/m_Q$ or by the strong-interaction coupling constant evaluated at a heavy-quark mass scale.

In order to gain further information about the $q^2$ behavior of heavy-to-light transition form factors, much more precise experimental data on exclusive semileptonic B- and D-meson decays are urgently needed.

\section*{Acknowledgments}
The work is supported by National Natural Science Foundation under contract
Nos.10675039 and 10735080. We thank Xinqiang Li for many helpful discussions.

\end{document}